\documentclass{article}
\usepackage{spconf,amsmath,graphicx,hyperref}

\usepackage{comment}
\usepackage{multirow}
\usepackage{booktabs}
\usepackage{color}
\usepackage{subcaption}
\usepackage{amssymb}

\title{
ProLAP: Probabilistic Language-Audio Pre-Training
}
\name{
Toranosuke Manabe$^{1,2,\ast}$,
Yuchi Ishikawa$^{1}$,
Hokuto Munakata$^{1}$,
Tatsuya Komatsu$^{1}$
\thanks{$\ast$ This work was done during an internship at LY Corporation.}
}
\address{
$^1$LY Corporation,
$^2$Keio University
}
\begin{document}
\ninept
\maketitle
%

\begin{abstract}
Language-audio joint representation learning frameworks typically depend on deterministic embeddings,
assuming a one-to-one correspondence between audio and text.
In real-world settings, however, the language-audio relationship is inherently many-to-many:
one audio segment can be described by multiple captions and vice versa.
To address this, we propose Probabilistic Language-Audio Pre-training (ProLAP),
which models multiplicity as the spread of probability distributions
in a joint language-audio embedding space.
To train the intra-modal hierarchical relationship effectively,
we also introduce two objectives:
(i) hierarchical inclusion loss to promote semantic hierarchical understanding of inputs
and (ii) mask repulsive loss to improve the efficiency of learning when optimizing the hierarchical inclusion loss.
With this training strategy, our model can learn the hierarchical structure inherent in the data even from small datasets, in contrast to prior probabilistic approaches that rely on large-scale datasets.
In our experiments, ProLAP outperforms existing deterministic approaches on audio-text retrieval tasks.
Moreover, through experiments on the audio traversal task introduced in this paper,
we demonstrate that ProLAP captures the plausible semantic hierarchy.
%
\end{abstract}


\vspace{-2mm}
\section{Introduction}
\vspace{-1mm}

Learning joint representations of language and audio
has emerged as a powerful paradigm for open-vocabulary audio understanding.
Contrastive Language-Audio Pre-training (CLAP)~\cite{elizalde2023clap}
aligns audio clips and textual descriptions in a shared embedding space,
enabling zero-shot retrieval and classification in diverse acoustic domains.

Most language–audio models~\cite{10095969, 10448504} implicitly
assume a one-to-one correspondence between audio and text
and map each input to a single deterministic point in the embedding space.
In real-world settings, however, the language-audio relationship is inherently many-to-many and involves uncertainty:
a single audio segment can be described by multiple valid captions
at different levels of specificity
(e.g., string instrument $\to$ guitar $\to$ acoustic guitar),
and multiple paraphrases at the same level (e.g. ``Men talking followed by an engine starting'' and ``An engine starts after men talk'').

In computer vision, numerous works address many-to-many correspondences between language and images.~\cite{desai2023meru, alper2024hierarcaps, chun2021pcme, chun2024improved, chun2024probabilistic}
Within this line of work, learning frameworks based on hierarchical captions~\cite{alper2024hierarcaps}
and contrastive learning with probabilistic representations~\cite{chun2021pcme, chun2024improved}
have improved performance on downstream tasks.

Notably, ProLIP~\cite{chun2024probabilistic} models
the uncertainty of inputs and successfully learns the semantic hierarchy without hierarchical captions,
using probabilistic representations.
The key idea is to treat the uncertainty as the variance of a representation.
To facilitate learning the semantic hierarchy, ProLIP introduces inclusion loss.
This loss models the hierarchy as inclusion relationships
among probabilistic representations:
text descriptions are treated as more uncertain than their corresponding images,
and masked inputs as more uncertain than their unmasked counterparts.
While probabilistic representations have been successful in computer vision,
their success raises the question of
whether these approaches are effective for joint language-audio learning
and what modifications are required.

We propose Probabilistic Language-Audio Pre-training (ProLAP), a probabilistic extension of CLAP~\cite{elizalde2023clap}.
ProLAP represents each input (audio or text) as a probability distribution over 
the joint embedding space rather than as a deterministic vector.
However, we find empirically that a straightforward probabilistic extension
is insufficient to capture the semantic hierarchy in audio and text.
Therefore, we introduce hierarchical inclusion loss and mask repulsive loss to promote hierarchical learning.
With this training strategy, ProLAP explicitly captures
(i) multimodal semantics arising from many-to-many audio–text correspondences
and (ii) hierarchical relations between general and specific concepts.

In the experiments, ProLAP outperforms standard CLAP and CLAP with sigmoid loss~\cite{zhai2023sigmoid}
on audio-text retrieval tasks.
Additionally, to evaluate semantic hierarchical understanding, we introduce audio traversal task, analogous to the image traversal task~\cite{desai2023meru}.
The audio traversal task traces a straight line 
in embedding space from an audio embedding to the abstract anchor,
retrieving the nearest text embedding at sampled points.
The results indicate that ProLAP better captures the semantic hierarchy than baselines.


\begin{figure*}[t]
    \centering
    \small
    \includegraphics[width=0.95\linewidth]{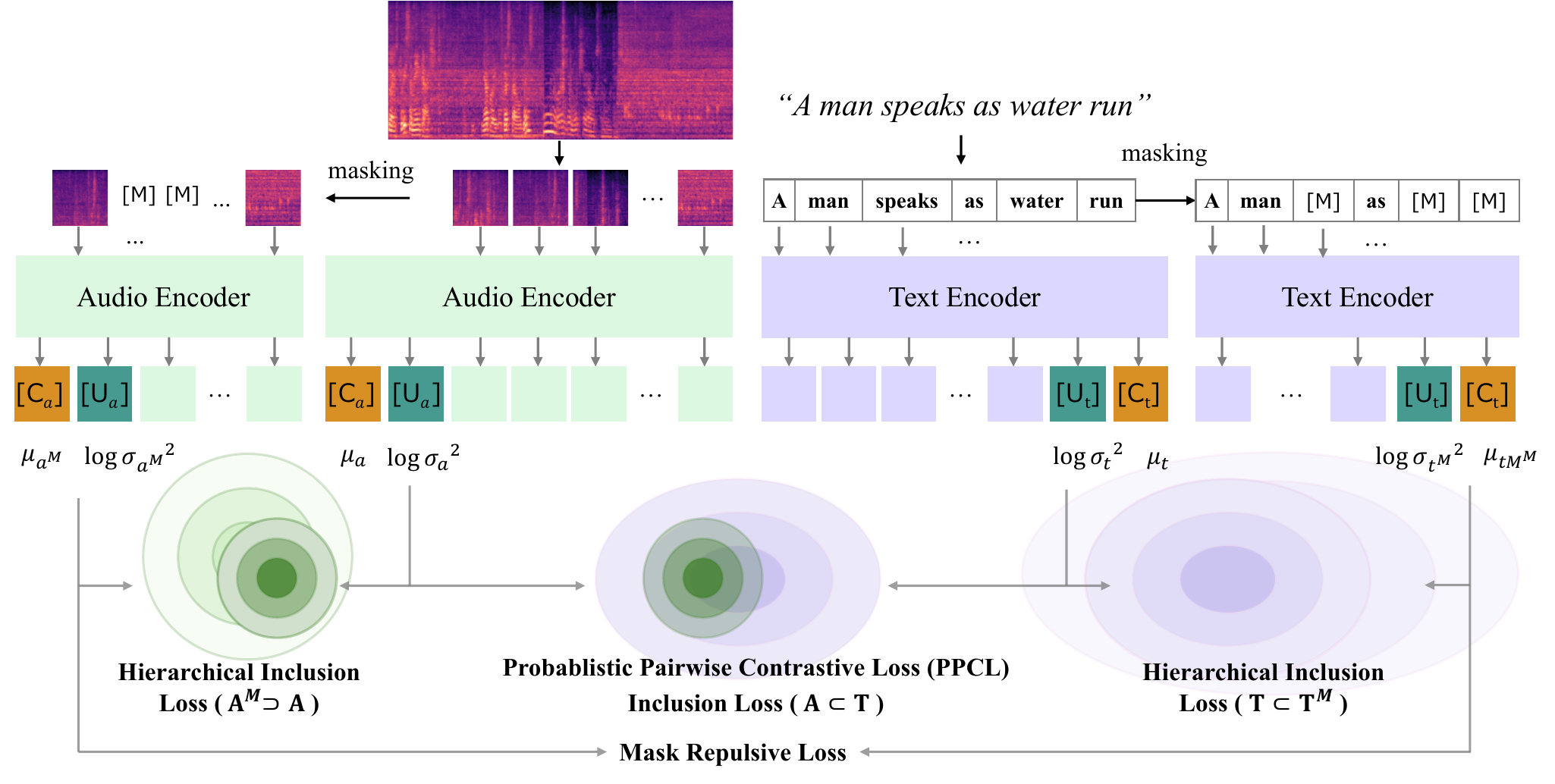}
    \vspace{-3mm}
    \caption{
    \textbf{Overview of ProLAP. }
    ProLAP models the representation as a Gaussian random variable following $\mathcal{N}(\mu, \sigma^2)$.
    Audio-text alignment is learned using PPCL (Eq.~\ref{eq:PPCL}), while semantic hierarchical understanding is encouraged by inclusion loss (Eqs.~\ref{eq:inc} and \ref{eq:h-inc}).
    Mask repulsive loss (Eq.~\ref{eq:MR}) prevents degenerate of representations for masked inputs.
    }
    \label{fig:top}
\end{figure*}

\vspace{-2mm}
\section{Proposed Method}
\vspace{-1mm}

In this paper, we propose Probabilistic Language-Audio Pre-training (ProLAP), 
which learns the joint language-audio representations
with probabilistic modeling of inputs.
Building on ProLIP~\cite{chun2024probabilistic},
we follow its probabilistic formulation,
and our baseline ProLAP is a straightforward application of the ProLIP for audio-text data.
We also propose two loss functions to promote hierarchical learning:
hierarchical inclusion loss and mask repulsive loss.
In the following sections, we first provide an overview of ProLAP,
and then we explain the two loss objectives used in ProLIP to learn the cross-modal probabilistic representations.
Finally, we introduce two additional loss functions to promote semantic hierarchy understanding.

\vspace{-2mm}
\subsection{Model Overview}
As shown in Fig.~\ref{fig:top}, ProLAP consists of audio and text encoders,
and aims to learn joint language-audio representations.
We model each input as a Gaussian random variable with diagonal covariance, parameterized by a mean vector $\mu$ and a variance vector $\sigma^{2}$.
\vspace{-2mm}
\subsection{ProLAP Baseline Objective}
\subsubsection{Probabilistic Pairwise Contrastive Loss}
To incorporate the distance between the probabilistic representations into contrastive learning,
we use the probabilistic pairwise contrastive loss (PPCL) following ProLIP~\cite{chun2024probabilistic}.
We replace cosine similarity with the corrected similarity $s(\cdot)$,
which minimizes the closed-form sampled distance (CSD)~\cite{chun2024improved}:
\vspace{-1mm}
{
\small
\begin{equation}
\label{eq:similarity}
    s(Z_a, Z_t) = \mu_a^\mathsf{T}\mu_t -\frac{1}{2}tr(\Sigma_a+\Sigma_t),
\end{equation}
}
\vspace{-1mm}

\noindent where $Z_a\sim \mathcal{N}(\mu_a, \Sigma_a)$ and $Z_t\sim \mathcal{N}(\mu_t, \Sigma_t)$ are Gaussian random variables corresponding to audio and text.
PPCL is formulated as the sigmoid loss~\cite{Zhai_2023_ICCV} by substituting the corrected similarity $s(\cdot)$:
\vspace{-1mm}
{
\small
\begin{equation}
\label{eq:PPCL}
    \mathcal{L}_{\mathrm{PPCL}}(Z_a, Z_t) = -\log\frac{1}{1 + \exp(y_{at}(-\alpha s(Z_a, Z_t) + \beta)},
\end{equation}
}
\vspace{-1mm}

\noindent where $\alpha$ and $\beta$ are learnable scalars, and $y_{at}$ equals $+1$ if audio $a$ and text $t$ are a matched pair, and $-1$ otherwise.
\vspace{-1mm}
\subsubsection{Inclusion Loss}
To model the semantic hierarchy, we use the inclusion loss following ProLIP~\cite{chun2024probabilistic}.
First, we define the following inclusion score (test statistic) to quantify the asymmetric inclusion $Z_1\subset Z_2$:
\vspace{-1mm}
{
\small
\begin{equation}
\label{eq:inclusion_test}
\begin{aligned}
    \mathcal{H}(Z_1\subset Z_2) = &\log\int_{-\infty}^{\infty}p_1^2(x)p_2(x)dx -\log\int_{-\infty}^{\infty}p_1(x)p^2_2(x)dx,
\end{aligned}
\end{equation}
}

\noindent where $p_1(x)$ and $p_2(x)$ are the probability density functions (pdf) corresponding to $Z_1$ and $Z_2$, respectively.
We then map this score to a probability via a logistic link
and use its negative log-likelihood as the inclusion loss,
which encourages inclusion of $Z_1$ in $Z_2$:
\vspace{-1mm}
{
\small
\begin{equation}
\label{eq:inc}
    \mathcal{L}_{\mathrm{inc}}(Z_1\subset Z_2) = -\log\frac{1}{1 + \exp(-c\mathcal{H}(Z_1\subset Z_2))},
\end{equation}
}

\noindent where $c$ is a positive scalar constant.
We designed ProLAP to learn two types of inclusion relationships: cross-modal and intra-modal (as shown in Fig.~\ref{fig:top}).
For the cross-modal case, we assume that text captions is more uncertain
than its audio segment.
We encode this relationship with the inclusion loss $\mathcal{L}_{\mathrm{inc}}(Z_a\subset Z_t)$.
For the intra-modal case, we assume that a masked input is
considered more uncertain than its raw counterpart.
We encode this relationship with the inclusion loss $\mathcal{L}_{\mathrm{inc}}(Z_{p}\subset Z_{p^M})$, where $p$ indexes the modality (audio $a$ or text $t$) and $p^{M}$ denotes the masked input.
\vspace{-2mm}
\subsection{Hierarchical Understanding}
Empirically, we find the ProLAP baseline
is insufficient for learning the semantic hierarchy in audio and text
(see details in Sec.~\ref{sec:uncertainty_compare}).
To address this, we introduce two loss functions to promote semantic hierarchical understanding.
\vspace{-2mm}
\subsubsection{Hierarchical Inclusion Loss}\label{sec:hierarchical-inclusion}
We extend intra-modal inclusion loss by introducing hierarchical masks to promote more fine-grained hierarchical understanding.
We define a hierarchy of binary masks $\{M_i\}_{i=0}^{L}$ recursively using the Hadamard product:
{
\vspace{-1mm}
\small
\begin{equation}
M_i \;=\; M_{i-1} \odot R_{i-1},\qquad i=1,\dots,L-1,
\end{equation}
\vspace{-1mm}
}

\noindent where each $R_i \in \{0,1\}^{d}$ is a random mask, and $M_0$ and $M_L$ represent the full mask and the zero mask, respectively.
The hierarchical inclusion loss in ProLAP is defined as follows:
\vspace{-1mm}
\begin{equation}
\label{eq:h-inc}
\mathcal{L}^h_{\mathrm{inc}}(Z_{p}) = 
\sum_{i=0}^{L - 1}\mathcal{L}_{\mathrm{inc}}(Z_{p^{M_{i+1}}} \subset Z_{p^{M_i}}),
\end{equation}
\vspace{-1mm}
\noindent where $Z_p$ is the Gaussian random variable, and $Z_{p^{M_i}}$ denotes 
$Z_p$ masked by $M_i$.
\vspace{-2mm}
\subsubsection{Mask Repulsive Loss}
Moreover, our ProLAP baseline produces degenerate representations for masked inputs
(See Fig.~\ref{fig:audio_visualization_baseline}).
If the model cannot adequately distinguish the representations of masked inputs,
it fails to faithfully encode the unmasked content, thereby undermining
hierarchical learning with the mask-based inclusion loss described above.
Therefore, we introduce mask repulsive loss, 
which pushes representations for masked inputs away from each other. 
Mask repulsive loss is defined as a negative-only PPCL:
{
\small
\begin{equation}
\vspace{-1mm}
\label{eq:MR}
\mathcal{L}_{\mathrm{MR}}(Z_{p}, Z_{q}) = -\sum_{i=1}^{L-1}\log\frac{1}{1 + \exp(y_{pq}(-\alpha s(Z_{p^{M_i}}, Z_{q^{M_i}})) + \beta)},
\end{equation}
}

\noindent
where $Z_p$ and $Z_q$ are Gaussian random variables from the same modality,
and $y_{pq}$ equals 0 if $p$ and $q$ refer to the same sample, and $-1$ otherwise.
We stop gradients with respect to the uncertainty parameters in mask repulsive loss,
because increasing uncertainty can trivially decrease the loss 
without encouraging distinguishable representations. 
In sum, the intra-modal loss is defined as follows:
{
\small
\begin{equation}\label{eq:intra-modal-loss}
\begin{aligned}
&\mathcal{L}_{\mathrm{intra}}(Z_p)
= \\
&\sum_{Z_p\in\mathcal{P}} \left[
\lambda_1\mathcal{L}^h_{\mathrm{inc}}(Z_p)
+\lambda_2\sum_{Z'_{p}\in\mathcal{P}}\mathcal{L}_{\mathrm{MR}}(Z_p, Z'_{p}) +\gamma\mathcal{L}_{\mathrm{VIB}}(Z_p)
\right],
\end{aligned}
\end{equation}
}

\noindent where $\lambda_1$, $\lambda_2$, and $\gamma$ are weighting coefficients
for the respective loss terms,
and $\mathcal{L}_{\mathrm{VIB}}$ is a regularizer applied to each
Gaussian embedding to prevent variances from collapsing
~\cite{chun2021pcme,chun2024improved}.
Eq.~\ref{eq:PPCL}, \ref{eq:inc}, and \ref{eq:intra-modal-loss} all together,
we obtain the following learning objective:
{
\small
\begin{equation}\label{eq:total}
\begin{aligned}
\mathcal{L} = 
&\sum_{Z_a\in \mathcal{A}}\sum_{Z_t\in \mathcal{T}}\mathcal{L}_{\mathrm{PPCL}}(Z_a, Z_t) 
+\sum_{Z_a\in\mathcal{A}} \mathcal{L}_{\mathrm{intra}}(Z_a) \\
&+\sum_{Z_t\in\mathcal{T}} \mathcal{L}_{\mathrm{intra}}(Z_t)
+\sum_{(Z_a, Z_t)\in(\mathcal{A}, \mathcal{T})}\lambda_3 \mathcal{L}_{\mathrm{inc}}(Z_a \subset Z_t), 
\end{aligned}
\end{equation}
}

\noindent where $\mathcal{A}$ and $\mathcal{T}$ denote sets of audio and text data,
respectively, and $\lambda_3$ is the weighting coefficient for the cross-modal inclusion loss.

\vspace{-1mm}
\section{Experiments}
\vspace{-1mm}
\begin{table*}[]

\centering
\caption{\textbf{Results on audio-text retrieval tasks using AudioCaps (AC) and ClothoV2 (CL).}}
\vspace{-3mm}

\begin{tabular}{lccccccccc}

\toprule[1.2pt]
\multicolumn{10}{c}{\textbf{AudioCaps Dataset}} \\
\toprule[1.2pt]
\multirow{2}{*}{\textbf{Method}} & \textbf{Training} & \multicolumn{4}{c}{\textbf{Text-to-Audio Retrieval}} & \multicolumn{4}{c}{\textbf{Audio-to-Text Retrieval}} \\
                  & \textbf{Dataset}             & \textbf{R@1} & \textbf{R@5} & \textbf{R@10} & \textbf{mAP@10} & \textbf{R@1} & \textbf{R@5} & \textbf{R@10} & \textbf{mAP@10} \\
\midrule[0.5pt]
CLAP (InfoNCE)    &       AC          & 41.90 & \textbf{77.24} & \textbf{89.24} & 57.30 & 39.75 & \textbf{78.32} & \textbf{89.35} & 55.84 \\
CLAP (SigLIP)     &       AC          & 41.45 & 76.22 & 88.67 & 56.22 & 40.88 & 76.90 & 88.34 & 56.22 \\
ProLAP (Ours)     &       AC          & \textbf{43.15} & 77.01 & 89.01 & \textbf{57.65} & \textbf{41.79} & 77.12 & 88.45 & 56.52 \\
ProLAP w/ $\mathcal{L}^h_{\mathrm{inc}}, \mathcal{L}_{\mathrm{MR}}$ (Ours)   &       AC          & 42.36 & 76.56 & 88.22 & 57.23 & \textbf{41.79} & 78.48 & 88.67 & \textbf{56.72} \\
\midrule[0.5pt]
CLAP (InfoNCE)    &       CL          & 27.63 & 62.85 & 76.78 & 42.45 & \textbf{26.73} & 57.19 & 71.57 & 39.43 \\
CLAP (SigLIP)     &       CL          & 27.97 & 62.40 & 77.01 & 42.49 & 24.46 & 59.23 & 76.10 & 39.20 \\
ProLAP (Ours)     &       CL          & 28.65 & \textbf{63.65} & 77.46 & 43.54 & 25.71 & 60.93 & 76.44& 40.56 \\
ProLAP w/ $\mathcal{L}^h_{\mathrm{inc}}, \mathcal{L}_{\mathrm{MR}}$(Ours)   
                  &       CL          & \textbf{28.77} & 63.53 & \textbf{78.26} & \textbf{43.72} & 26.05 & \textbf{61.27} & \textbf{77.01} & \textbf{40.99} \\
\midrule[1.2pt]
\midrule[1.2pt]
\multicolumn{10}{c}{\textbf{ClothoV2 Dataset}} \\
\midrule[1.2pt]
\multirow{2}{*}{\textbf{Method}} & \textbf{Training} & \multicolumn{4}{c}{\textbf{Text-to-Audio Retrieval}} & \multicolumn{4}{c}{\textbf{Audio-to-Text Retrieval}} \\
                  & \textbf{Dataset}             & \textbf{R@1} & \textbf{R@5} & \textbf{R@10} & \textbf{mAP@10} & \textbf{R@1} & \textbf{R@5} & \textbf{R@10} & \textbf{mAP@10} \\
\midrule[0.5pt]
CLAP (InfoNCE)    & AC & 20.00 & 44.98 & 59.23 & 31.03 & 17.22 & 44.69 & 57.42 & 28.63 \\
CLAP (SigLIP)     & AC & 19.14 & 45.07 & 58.66 & 30.13 & 17.99 & 44.31 & \textbf{59.52} & 29.19 \\
ProLAP (Ours)     & AC & 20.48 & 45.65 & \textbf{59.81} & 31.46 & \textbf{18.95} & 44.40 & 58.66 & \textbf{29.77} \\
ProLAP w/ $\mathcal{L}^h_{\mathrm{inc}}, \mathcal{L}_{\mathrm{MR}}$ (Ours)   & AC & \textbf{20.57} & \textbf{45.74} & 59.33 & \textbf{31.53} & 18.18 & \textbf{45.17} & 58.28 & 29.30 \\
\midrule[0.5pt]
CLAP (InfoNCE)    & CL & 20.67 & 46.12 & 59.43 & 31.72 & 19.43 & 45.65 & 59.71 & 30.62 \\
CLAP (SigLIP)     & CL & 18.85 & 45.07 & 57.80 & 29.90 & 17.89 & 45.26 & 59.71 & 29.85 \\
ProLAP (Ours)     & CL & 20.38 & \textbf{46.99} & \textbf{60.29} & 31.92 & 19.14 & \textbf{48.23} & 62.20 & 31.53 \\
ProLAP w/ $\mathcal{L}^h_{\mathrm{inc}}, \mathcal{L}_{\mathrm{MR}}$ (Ours)   & CL & \textbf{20.86} & 46.41 & 60.00 & \textbf{32.17} & \textbf{19.90} & 47.75 & \textbf{62.68} & \textbf{31.99} \\
\bottomrule[1.2pt]
\vspace{-3mm}
\end{tabular}
\label{tab:retrieval}
\end{table*}

\subsection{Experimental Setup}

To demonstrate the effectiveness of ProLAP,
we evaluate it on text-to-audio and audio-to-text retrieval tasks.
In our experiments, we first fine-tune the models
initialized with pre-trained CLAP~\cite{elizalde2023clap} weights,
and then evaluate the models on the retrieval tasks.
As baselines, we also train CLAP with the InfoNCE and
CLAP with the sigmoid loss proposed in SigLIP~\cite{zhai2023sigmoid}.
We use the similarity $s(\cdot)$ defined in Eq.~\ref{eq:similarity} 
when training and evaluating ProLAP.
Following prior works~\cite{Munakata2024,primus2024knowledge},
we report Recall@K (R@K) for $K \in \{1,5,10\}$
and mean average precision among the top 10 results (mAP@10).

\noindent\textbf{Datasets.}
We use three benchmarks for audio-text retrieval:
AudioCaps (AC)~\cite{kim2019audiocaps}]
and ClothoV2 (CL)~\cite{drossos2020clotho}.
AudioCaps contains 51,308 audio clips, each paired with a single caption.
ClothoV2 comprises 5,930 audio clips and has five captions per audio clip.

\noindent \textbf{Implementation Details.}
Our experimental settings is mostly aligned with that of~\cite{Munakata2024,komatsu2025leveraging}.
We train for 50 epochs with a batch size of 256.
For training, we use
Adam~\cite{DBLP:journals/corr/KingmaB14} as the optimizer.
The learning rate is scheduled with 
a cosine scheduler and one warm-up epoch.
The maximum learning rate is $1\times10^{-5}$.
We set $\lambda_1=5\times10^{-3}$, $\lambda_2=1\times10^{-4}$, 
$\lambda_3=5\times10^{-7}$, and $\gamma=1\times10^{-5}$ for all experiments.
We use $L=3$ for hierarchical inclusion loss.
We apply 75\% masking to the first 12.5\% of each batch.

\noindent \textbf{Architectures. }
Following the official implementation of CLAP\footnote{https://github.com/microsoft/CLAP},
we use HTS-AT~\cite{chen2022hts} as the audio encoder
and GPT-2~\cite{radford2019language} as the text encoder.
We use the end-of-sentence token as the \verb|[CLS]| token to estimate the mean,
and we add the \verb|[UNC]| token to estimate the variance in the text encoder.
Because HTS-AT is based on Swin Transformer~\cite{liu2021swin},
we cannot use token-level prediction and token dropping.
Instead, we add an additional head for uncertainty and
we use a learnable \verb|[MASK]| token for the audio encoder.

\vspace{-2mm}
\subsection{Comparison on AudioCaps and ClothoV2}

Table~\ref{tab:retrieval} compares the performance of models
trained either on AudioCaps or on ClothoV2.
Overall, ProLAP outperforms existing methods on both datasets.
Notably, the improvements are more pronounced on datasets not used for training,
suggesting that ProLAP is more robust to out-of-domain data than the baselines.

\vspace{-2mm}
\subsection{Understanding Learned Uncertainty}\label{sec:uncertainty_compare}

ProLAP is trained to estimate predictive uncertainty for the text and audio modalities.
To assess the quality of these estimates, we conduct two experiments.

\begin{figure}[t]
  \centering
  \centering
  \begin{subfigure}{0.24\textwidth}
    \centering
    \includegraphics[width=\linewidth]{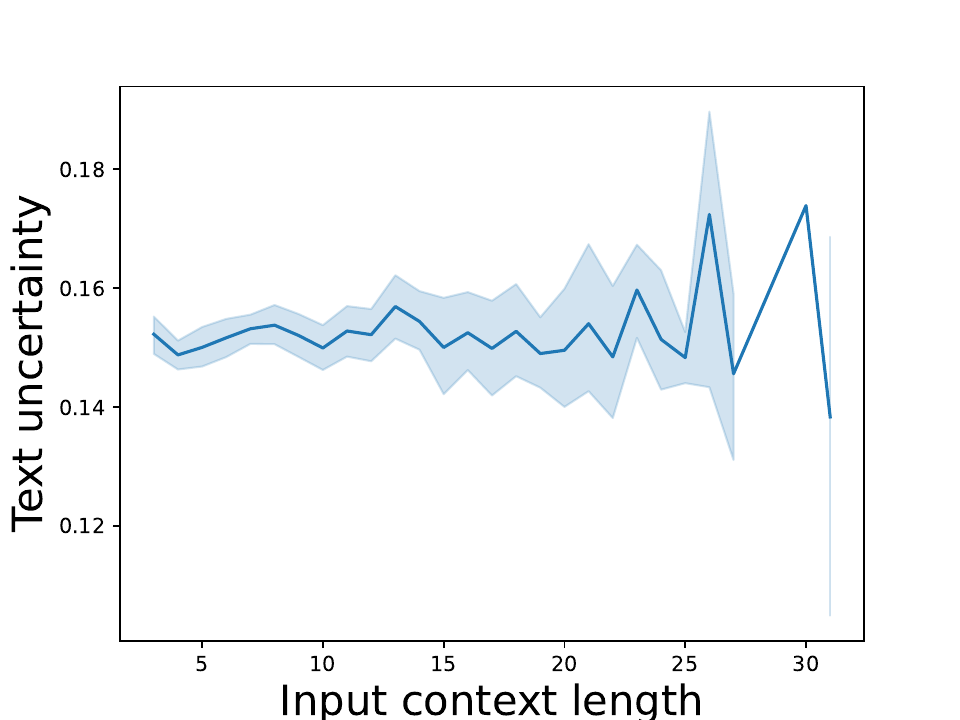}
    \caption{Baseline}
    \label{fig:uncertainty-audio}
  \end{subfigure}\hfill%
  \begin{subfigure}{0.24\textwidth}
    \centering
    \includegraphics[width=\linewidth]{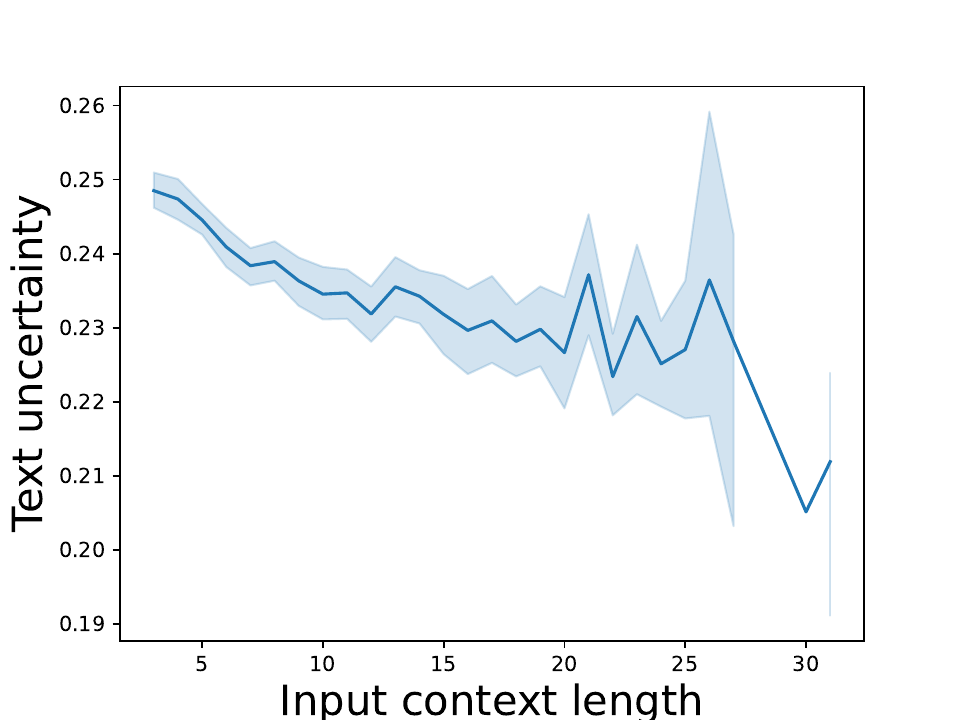}
    \caption{Proposed}
    \label{fig:uncertainty-text}
  \end{subfigure}
  \vspace{-7mm}
  \caption{\textbf{Text length vs uncertainty.}}
  \vspace{-3mm}
  \label{fig:text-length-vs-uncertainty}
\end{figure}
\noindent\textbf{Input length vs. uncertainty. }
We first examine the relationship between input text length
and predicted uncertainty on the AudioCaps dataset.
Intuitively, longer captions tend to be more specific (i.e., convey lower uncertainty).
As shown in Fig.~\ref{fig:text-length-vs-uncertainty},
the baseline predicts nearly constant uncertainty across text lengths, 
indicating that it fails to capture text uncertainty.
One plausible factor is data characteristics: compared with AudioCaps,
the DataComp~\cite{gadre2023datacomp} corpus used by ProLIP
differs markedly in both size (1.28B vs. 46K examples)
and caption length (1–75 vs. 2–30 tokens).
In contrast, when we apply our proposed loss,
we observe a downward trend in uncertainty as text length increases. 
These results suggest that the proposed method enables more data-efficient
uncertainty estimation, even with limited training data.

\begin{figure}[t]
  \centering
  \centering
  \begin{subfigure}{0.24\textwidth}
    \centering
    \includegraphics[width=\linewidth]{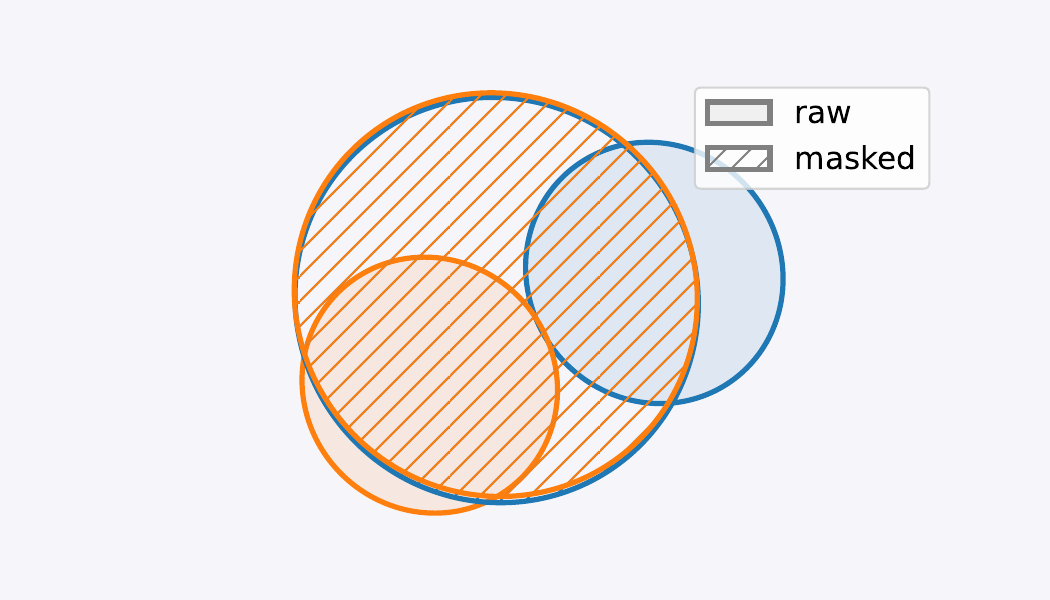}
    \caption{Baseline}
    \label{fig:audio_visualization_baseline}
  \end{subfigure}\hfill%
  \begin{subfigure}{0.24\textwidth}
    \centering
    \includegraphics[width=\linewidth]{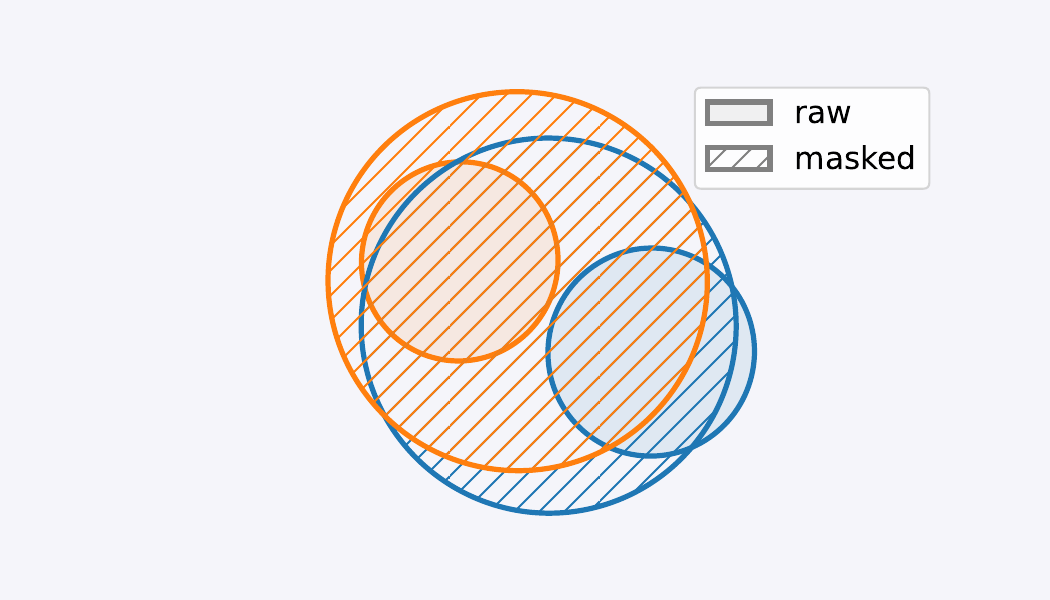}
    \caption{Proposed}
    \label{fig:uncertainty-text}
  \end{subfigure}
  \vspace{-6mm}
  \caption{
    \textbf{Visualization of probabilistic audio embeddings.}
  }
  \label{fig:vis-uncertainty}
\vspace{-1mm}
\end{figure}
\noindent\textbf{Qualitative Evaluation. }
Fig.~\ref{fig:vis-uncertainty} shows a visualization
of probabilistic audio embeddings of ProLAP.
Our baseline (ProLAP without the proposed losses) fails to distinguish inputs with different masks.
In contrast, ProLAP with the proposed loss functions produces distinguishable embeddings
for these inputs.
Moreover, it yields more appropriate inclusion relationships among the resulting distributions.

\vspace{-2mm}
\subsection{Audio Traversal}

\begin{table}[]
\centering
\vspace{-4mm}
\small
\caption{
\textbf{Audio traversal.}
R@1$^\text{Lv.1}$ indicates R@1 of the most abstract caption in HierarAudioCaps.
}
\vspace{-3mm}
\begin{tabular}{llll}
\toprule[1.2pt]
Method & Prec.   & R@1   & R@1$^\text{Lv.1}$ \\
\midrule[0.5pt]
CLAP   (InfoNCE)   &  12.77  &  13.46  &  8.48  \\
CLAP   (SigLIP)    &  10.60  &  10.66  &  8.48  \\
\midrule[0.5pt]
ProLAP (Ours) &  19.98  &  13.74  &  9.39  \\
ProLAP w/ $\mathcal{L}^h_{\mathrm{inc}}, \mathcal{L}_{\mathrm{MR}}$ (Ours)
         &  \textbf{27.33}  &  \textbf{15.16}  &  \textbf{11.76}  \\
\bottomrule[1.2pt]
\vspace{-5mm}
\end{tabular}
\label{tab:traversal}
\end{table}

To assess the hierarchy learned by ProLAP, we introduce audio traversals.
Analogous to image traversals~\cite{desai2023meru},
an audio traversal traces a straight line
in the embedding space from an audio embedding
to the most abstract anchor, \verb|[ROOT]|,
and retrieves the nearest text embedding at each of a sequence of points along this path.
Audio traversals also have practical value:
they can flag audio–text examples that show mismatches in abstraction level,
thereby supporting dataset curation or other downstream applications.

To evaluate audio traversal quantitatively, we introduce HierarAudioCaps,
which includes four levels of abstract captions
for each AudioCaps clip.
Level 4 denotes the original caption, while Level 1 denotes the most abstract caption.
Abstract captions are generated by gpt-oss~\cite{agarwal2025gpt}.

For deterministic models, \verb|[ROOT]| is the empty caption (\verb|""|).
For probabilistic models, following ProLIP~\cite{chun2024probabilistic},
\verb|[ROOT]| is defined as the average of the empty-caption embedding
and the most inclusive caption embedding for the query.
For evaluation, we use 50 equally spaced points
along the line segment between [ROOT] and the query embedding.
Table~\ref{tab:traversal} shows traversal performance on HierarAudioCaps.
The results imply that ProLAP has a better understanding of the semantic hierarchy,
and that our proposed losses contribute to improving the hierarchical representations.
\vspace{-2mm}
\subsection{Ablation Studies}
\begin{table}[]
\small
\vspace{-4mm}
\centering

\caption{
\textbf{Ablation study.}
Inclusion test indicates the proportion of cases in which Level 1 caption includes Level 4 counterpart.
}
\vspace{-3mm}
\begin{tabular}{cccccc}
\toprule[1.2pt]
\multirow{2}{*}{$\mathcal{L}^h_{\mathrm{inc}}$} & \multirow{2}{*}{$\mathcal{L}_{\mathrm{MR}}$}  & Retrieval & \multicolumn{2}{l}{Traversal} & Inclusion \\
 & & mAP@10 & Prec. & R@1     & test (\%) \\
\midrule[0.5pt]
           &            & \textbf{57.65} & 19.98 & 13.74 & 63.19 \\
\checkmark &            & 57.27          & 23.28 & \textbf{15.21} & 82.79 \\
           & \checkmark & 57.08          & 16.16 & 13.63 & 75.42 \\
\checkmark & \checkmark & 57.23          & \textbf{27.33} & 15.16 & \textbf{89.47} \\
\bottomrule[1.2pt]
\vspace{-5mm}
\end{tabular}
\label{tab:ablation}
\end{table}

To evaluate the effectiveness of the hierarchical inclusion loss 
and the mask-repulsive loss for learning hierarchical structure, 
we conduct an ablation study.
Table~\ref{tab:ablation} reports retrieval performance on AudioCaps and audio traversal performance on HierarAudioCaps,
The results indicate that the hierarchical inclusion loss 
contributes substantially to learning the hierarchy. 
By contrast, the mask repulsive loss alone slightly
degrades performance; however, when combined with the hierarchical inclusion loss, 
it has a complementary effect that further promotes hierarchical learning.

\vspace{-2mm}
\section{Conclusion}
\vspace{-1mm}
In this paper,
we propose Probabilistic Language–Audio Pre-training (ProLAP),
which integrates probabilistic representations into CLAP
to assess their effectiveness for language–audio learning.
To encourage hierarchical understanding,
we further introduce hierarchical inclusion loss and mask repulsive loss.
Experiments show that ProLAP outperforms CLAP on audio–text retrieval benchmarks
and likewise surpasses it on the audio traversal analysis,
thus validating the effectiveness of our loss design in capturing semantic hierarchy.

\clearpage
\bibliographystyle{IEEEbib}
\bibliography{main}

\end{document}